# Determination of the neutron capture cross sections of $^{232}$Th at 14.1 MeV and 14.8 MeV using the neutron activation method

Chang-Lin Lan(兰长林)[1,*] Bao-Lin Xie(解保林)[1] Meng Peng(彭猛)[1] Tao Lv(吕韬)[1]
Ze-en Yao(姚泽恩)[1] Jin-Gen Chen(陈金根)[2] Xiang-Zhong Kong(孔祥忠)[1]

[1] School of Nuclear Science and Technology, Lanzhou University, Lanzhou, Gansu 730000, China
[2] Shanghai Institute of Applied Physics, Chinese Academy of Sciences, Shanghai 201800, China

**Abstract**： The $^{232}$Th(n, γ)$^{233}$Th neutron capture reaction cross sections were measured at average neutron energies of 14.1 MeV and 14.8 MeV using the activation method. The neutron flux was determined using the monitor reaction $^{27}$Al(n,α)$^{24}$Na. The induced gamma-ray activities were measured using a low background gamma ray spectrometer equipped with a high resolution HPGe detector. The experimentally determined cross sections were compared with the literatures data, evaluated data of ENDF/B-VII, JENDL-4.0, and CENDL-3.1. The Excitation functions of $^{232}$Th(n,γ) reaction were also calculated theoretically using the TALYS 1.6 computer code.
**Keywords:** $^{232}$Th(n,γ)$^{231}$Th reaction, cross section, neutron activation method

## 1 Introduction

The database of cross sections for the reactions induced by neutrons around 14 MeV plays a key role for the applied physics and the design of nuclear reactors. Specifically, the neutron capture cross section of $^{232}$Th is an important parameter for the design of any nuclear reactor based on the Th--U fuel cycle[1]. In this cycle, $^{233}$U is the fissile isotope which is formed from $^{232}$Th by neutron capture followed by two β-decays.

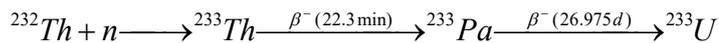
$$^{232}Th + n \longrightarrow {}^{233}Th \xrightarrow{\beta^-(22.3\min)} {}^{233}Pa \xrightarrow{\beta^-(26.975d)} {}^{233}U$$

Moreover, the production of the $^{233}$U depends on the $^{232}$Th(n,γ) reaction cross section is currently required within an accuracy of 1–2% in order to be used safely in simulated techniques for predicting the dynamical behavior of complex arrangements in the fast reactor or thorium molten salt reactors (TMSR)[2].

For the $^{232}$Th(n, γ)$^{233}$Th reaction, many researchers have been measured the cross section over a wide range of neutron energies from thermal to 17.28 MeV[3]. Around 14 MeV, only two data of the $^{232}$Th(n,γ) reaction cross-section are available at 14.5 MeV and 14.8 MeV using the activation technique, and there is relatively large disagreement and uncertainty between these data. Meanwhile, the discrepancies between the evaluated data files (ENDF/B-VII.1[4], JENDL-4.0[5], CENDL-3.1[6]) can reach 40% at the neutron energy of 14.1 MeV. Hence, the neutron capture reaction cross section of $^{232}$Th need further precise measurements to strengthen the reliability of the databases.

In this paper, we present the results of the $^{232}$Th(n, γ)$^{233}$Th reaction cross sections in neutron energies of 14.1 and 14.8 MeV using the activation method and off-line γ-ray spectrometric technique. The measured results are discussed and compared with the previous experimental data and evaluated data, as well as the calculated results of TALYS 1.6 code.

## 2 Experimental descriptions

The two targets (thorium dioxide powder of 99.7% purity) and monitor (aluminum foil of 99.99% purity) were made into circular disks with a diameter of 20 mm for irradiation. The thickness of ThO$_2$ samples with a thickness of 1.05 mm and 1.07 mm, and the thickness of Al samples were 0.06 mm. A stack of Al-Th-Al was made from these two samples and mounted at $0^0$ and $90^0$ angles relative to the deuteron beam direction.

The irradiations were carried out using the CPNG-600 neutron generator at China Institute of Atomic Energy (CIAE). The neutrons with the yield of about ~1.5×10$^{10}$ n/4π·s were produced by the T(d,n)$^4$He reaction. The ion beam current was up to 300 μA with effective deuteron energy of 300 keV. A solid tritium-titanium (T-Ti) target was used in the generator with the thickness of 1.0 mg/cm$^2$.

The neutron energies in these positions were calculated by the Q equation [7] and compared with the method of cross-section ratios for $^{90}$Zr(n,2n)$^{89m+g}$Zr and $^{93}$Nb(n,2n)$^{92m}$Nb reactions[8,9] before irradiation. The determined neutron energy was 14.1±0.2 MeV and 14.8±0.2 MeV, respectively.

During irradiation, the neutron flux was determined using the monitor reaction $^{27}$Al(n,α)$^{24}$Na, and the variation of the neutron yield was monitored by accompanying α-particles so that the corrections could be made for the fluctuation of neutron flux. The schematics view of the accompanying a-particle monitor was same as shown in Ref. [10].The Au–Si surface barrier detector used in 135° accompanying α-particle tube was at a distance of 110 cm from the target.

The off-line γ-ray measurements were performed by a coaxial GMX60 HPGe detector (ORTEC, made in USA) with a relative efficiency of 68% and an energy resolution of 1.82 keV FWHM at 1.33 MeV. The efficiency was calibrated with an standard $^{152}$Eu source previously. The uncertainty in the absolute efficiency curve at 50 mm was estimated to be ~3%.

## 3 Determination of $^{232}$Th(n,γ)$^{231}$Th reaction cross-sections

The decay parameters of the product radioisotopes used in the present work are taken from the Refs. [11, 12] and are summarized in Table 1.

Table 1. Reactions and associated decay data of activation products

| Reaction | Abundance of target isotope η (%) | Half-life of product $T_{1/2}$ | γ-ray energy $E_γ$ (keV) | γ-ray intensity $I_γ$ (%) |
|---|---|---|---|---|
| $^{232}$Th (n,γ) $^{233}$Th | 100 | 22.3 m | 86.48 | 2.76 |
| → $^{233}$Pa | | 26.975 d | 300.129 | 6.63 |
| | | | 311.904 | 38.5 |
| | | | 340.476 | 4.45 |
| $^{27}$Al(n, α) $^{24}$Na | 100 | 14.951 h | 1368.63 | 100 |

As shown in the Table.1, the half-life of the isotope $^{233}$Th is 22.3 minutes, which decays to $^{233}$Pa ($T_{1/2}$=26.975 d) within 3 hours. In view of this, the $^{232}$Th(n,γ)$^{231}$Th cross-section can be calculate from the observed photo-peak activity of $^{233}$Pa within the long cooled γ spectrum. In the calculation, the photo-peak activity of the 311.904 keV γ-line was selected instead of 300.129 keV and 340.476 keV. This is to avoid the interference of the 300.1 keV γ-line of $^{212}$Pb which is the α decay product from $^{232}$Th, and reduce the statistical error of 340.476 keV because of its low γ-ray intensity.

The following activation formula was used to determine the measured $^{232}$Th(n,γ)$^{231}$Th cross sections which proposed by X.Z. Kong et al.[13]

$$\sigma_x = \frac{[\eta \varepsilon I_\gamma mKSD]_m}{[\eta \varepsilon I_\gamma mKSD]_x} \frac{[\lambda FCA]_x}{[\lambda FCA]_m} \sigma_m \qquad (1)$$

where the footnotes $m$ and $x$ represent the terms of the monitor reaction and the measured reaction, respectively. $\sigma_m$ is the cross section of monitor reaction, $\varepsilon$ is the full-energy peak efficiency of the measured characteristic γ-ray, $I_\gamma$ is the γ-ray intensity, $\eta$ is the abundance of the target nuclide, $m$ is the mass of sample, $S = 1 - e^{-\lambda T}$ is the growth factor of the residual nuclide, $\lambda$ is the decay constant, and $T$ is the total irradiation time. $D = e^{-\lambda t_1} - e^{-\lambda t_2}$ is the counting collection factor, $t_1$ and $t_2$ are time intervals from the end of the irradiation to the start of counting and end of counting, respectively, $A$ is the atomic weight, $C$ is the measured full-energy peak area and $F$ is the total correction factor of the activity:

$$F = F_s \times F_c \times F_g \qquad (2)$$

where $F_s$, $F_c$ and $F_g$ are the correction factors for the self-absorption of the sample at a given γ-energy, the coincidence sum effect of cascade γ-rays in the investigated nuclide and in the counting geometry, respectively.

$K$ is the neutron fluency fluctuation factor,

$$K = [\sum_{i}^{L} \Phi_i (1 - e^{-\lambda \Delta t_i}) e^{-\lambda T_i}] / \Phi S \qquad (3)$$

In this calculation, we divided the total irradiation time into $L$ parts. where $L$ is the number of time intervals into which the irradiation time is divided, $\Delta T_i$ is the duration of the $i^{th}$ time interval, $T_i$ is the time interval from the end of the $i^{th}$ interval to the end of irradiation, $\Phi_i$ is the neutron flux averaged over the sample during the $\Delta T_i$, and $\Phi$ is the neutron flux averaged over the sample during the total irradiation time $T$.

**4 Nuclear model calculations**

In this work, the excitation function of the $^{232}$Th(n,γ)$^{231}$Th reaction cross sections at different neutron energies from threshold to 20 MeV was calculated theoretically using the computer code TALYS, version 1.6.[14, 15]. The TALYS-1.6 code system is able to analyze and predict nuclear reactions based on physics models and parameterizations. It calculate nuclear reactions involving neutrons, photons, protons, deuterons, tritons, $^3$He and alpha-particles, in the 1 keV - 200 MeV energy range and for target nuclides of mass 12 and heavier.

For the $^{232}$Th target, The optical model parameters for neutrons were obtained by a local potential proposed by Koning and Delaroche [16]. Similarly, the compound nucleus contribution was calculated by the Hauser– Feshbach model [17]. The two-component exciton model developed by Kalbach [18] was used for calculating the pre-equilibrium contribution. The default values were used for parameters concerning nuclear masses, ground-state deformations, discrete levels, decay schemes, and strength functions. Meanwhile, all the possible outgoing channels for a given neutron energy were considered including inelastic and fission channels.

## 5 Results and discussion

Cross sections for $^{232}$Th(n,γ)$^{231}$Th reaction at neutron energies of 14.1 and 14.8 MeV were determined. Corrections were also made for γ-ray self-absorption in the sample, for γ-ray coincidence summing effects, for fluctuation of the neutron flux during the irradiation and for sample geometry. The uncertainties in our results mainly include the counting statistics (10–15%), detector efficiency (~3%), neutron energy and fluency uncertainties (~2.5%), monitor reaction cross section (~1.5%), weight of samples (<0.1%), self-absorption of γ-ray (~1%), coincidence summing effect of cascade γ-rays (~1%), sample geometry (~1%), the irradiation, cooling and measuring times (<0.8%), etc. The overall uncertainty was carried out using the quadrature method.

The cross section measured in the present work and the values around neutron energy 14 MeV given in the literatures were summarized in Table 2 together with the cross section of the monitor reaction $^{27}$Al(n,α)$^{24}$Na. In the calculations, the cross

Submitted to 'Chinese Physics C'sections for the $^{27}$Al(n,α)$^{24}$Na monitor reaction were obtained by interpolating the evaluated values of the literature[12].

Table 2. Summary of the cross sections

| Reaction | Cross-sections (in mb) at various neutron energies (in MeV) | | |
| --- | --- | --- | --- |
| | $E_n$ = 14.1±0.2 MeV | $E_n$=14.5 MeV | $E_n$=14.8±0.2 MeV |
| $^{232}$Th(n,γ)$^{231}$Th | 2.320±0.378 | | 2.201±0.328 |
| | | 5.2±0.624 [20] | 1.569±0.257 [19] |
| $^{27}$Al(n,α)$^{24}$Na | 121.6±0.6 | | 111.9±0.5 |

It can be seen from Table 2, our measured results are in good agreement with the literature data [19] but much lower than [20]. For purpose of comparison, the $^{232}$Th(n,γ)$^{231}$Th reaction cross sections from the present work and the similar experimental data from the literatures[21-27] were plotted in Fig. 1 together with the evaluated data [4-6] as well as the theoretical values from computer code TALYS1.6 for the energy range of 1MeV to 20 MeV.

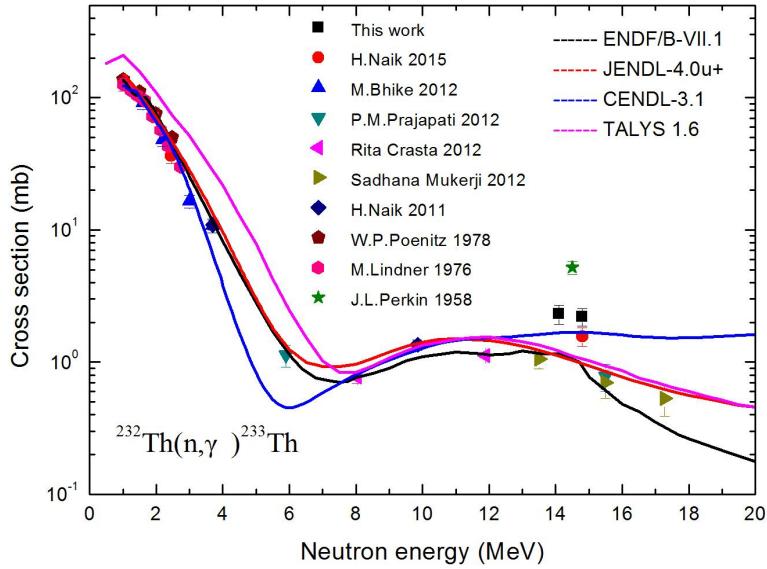

Fig. 1. Plot of the experimental, evaluated and theoretical calculated $^{232}$Th(n,γ)$^{231}$Th reaction cross sections as a function of the neutron energy from 1MeV to 20 MeV. Experimental values are in different symbols, whereas the evaluated and theoretical values from TALYS are in solid lines of different colors.

As shown in Fig.1, the $^{232}$Th(n,γ)$^{231}$Th reaction cross sections show a sharp decreasing trend from 1 MeV to the 7.5 MeV and then increasing up to the neutron energy of 12 MeV. This dip mainly due to the opening of the (n,2n) and (n,nf) reaction channels . It can be also seen from Fig.1 that the experimental cross sections of $^{232}$Th within neutron energy of 14.1- -14.8 MeV are significantly higher than the calculated results from TALYS and evaluated data from ENDf/B-VII.1, JENDL-4.0 except the evaluated data of CENDL-3.1.

## 6 Conclusion

We have measured the activation neutron capture cross sections of $^{232}$Th at neutron energy of 14.1 and 14.8 MeV using the off-line gamma ray spectrometric technique. During the work we used the lateast nuclear data for decay characteristics of the product radionuclides. The excitation function of the $^{232}$Th(n,γ)$^{231}$Th reaction were also claculated using computer code TALYS 1.6 and compared with the recent evaluated data of ENDF/B-VII.1, JENDL-4.0 and CENDL-3.1. Our results are useful for verifying the accuracy of nuclear models used in the calculation of cross sections and also be useful for the design, evaluation and construciton of nuclear reactor based on the $^{232}$Th - $^{233}$U fuel cycle.

**Acknowledgments**

We would like to thank the Intense Neutron Generator group at China Institute of Atomic Energy for performing the irradiations. This work is supported by the Chinese TMSR Strategic Pioneer Science and Technology Project-The Th-U Fuel Physics Term (No. XDA02010100) and the National Natural Science Foundation of China . (No. 11205076).